\documentclass[aps,preprint,showpacs,groupedaddress]{revtex4}
\usepackage {graphicx,epsfig,graphics,color}

\begin{document}

\title{Ferromagnetic resonance force microscopy on a thin permalloy film}

\author{E. Nazaretski}
\affiliation{Los Alamos National Laboratory, Los Alamos, NM
87545}
\author{D. V. Pelekhov}
\affiliation{Department of Physics, Ohio State University, Columbus OH 43210,}
\author{I. Martin}
\affiliation{Los Alamos National Laboratory, Los Alamos, NM
87545}
\author{M. Zalalutdinov}
\affiliation{SFA Inc., Crofton, MD 21114,}
\author{J. W. Baldwin}
\affiliation{Naval Research Laboratory, Washington DC 20375}
\author{T. Mewes}
\affiliation{Department of Physics and Astronomy, University of
Alabama, Tuscaloosa, AL 35487}
\author{B. Houston}
\affiliation{Naval Research Laboratory, Washington DC 20375}
\author{P. C. Hammel}
\affiliation{Department of Physics, Ohio State University, Columbus OH 43210,}
\author{R. Movshovich}
\affiliation{Los Alamos National Laboratory, Los Alamos, NM
87545}


\begin{abstract}
Ferromagnetic Resonance Force Microscopy (FMRFM) offers a means of performing
local ferromagnetic resonance. We have studied the evolution of the FMRFM force
spectra in a continuous 50 nm thick permalloy film as a function of probe-film
distance and performed numerical simulations of the intensity of the FMRFM
probe-film interaction force, accounting for the presence of the localized
strongly nonuniform magnetic field of the FMRFM probe magnet. Excellent
agreement between the experimental data and the simulation results provides
insight into the mechanism of FMR mode excitation in an FMRFM experiment.
\end{abstract}

 \maketitle

Magnetic resonance force microscopy (MRFM) offers a very high
sensitivity approach to detection of magnetic resonance and has
demonstrated three dimensional imaging with excellent spatial
resolution. Proposed by J.A. Sidles \cite{Sidles 1991}, it has
been used for the detection of electron spin resonance (ESR)
\cite{Rugar 1992}, nuclear magnetic resonance (NMR) \cite{Rugar
1994}; recently Rugar and co-workers reported detection of a force
signal originating from a single electron spin \cite{Rugar 2004},
emphatically demonstrating MRFM sensitivity. Incorporating basic
elements of MRI, MRFM can provide much higher spatial resolution
than conventional MRI. Electron spin density images with
micrometer scale resolution in an arbitrarily shaped sample can be
deconvolved from MRFM spatial force maps \cite{Zuger 1993}. MRFM
image deconvolution requires a thorough understanding of the
underlying interaction between the MRFM probe and the object
imaged. This deconvolution process, analyzed for the case of
noninteracting spins in a paramagnetic sample is given in Ref.
\onlinecite{Suter 2002}. Recently, ferromagnetic resonance (FMR)
has been detected by MRFM in YIG bar \cite{Zhang 1996}, YIG dot
\cite{Naletov 2003, Charbois 2002}, YIG film \cite{Urban 2006} and
permalloy dots \cite{Mewis 2006}.  The role of the probe magnet in
FMRFM is dual: it both perturbs the FMR modes, and detects the
force signal. However, in Ref. \cite{Zhang 1996, Naletov 2003,
Charbois 2002} the FMR modes were only weakly modified by the tip
field.

In this letter we focus on the regime when the effect of the tip
is non-perturbative: the field inhomogeneity due to the tip field
strongly modifies the resonance modes as well as leads to the
formation of the {\em local} resonance under the tip. We report
FMRFM spectra from a continuous, 50 nm thick permalloy film,
performed with a cantilever with a nearly spherical micron-size
magnetic tip. We report the evolution of the FMRFM spectra as a
function of the tip-sample spacing, and propose a model which
describes the observed behavior.

\begin{figure}
\includegraphics [angle=0,width=7cm]{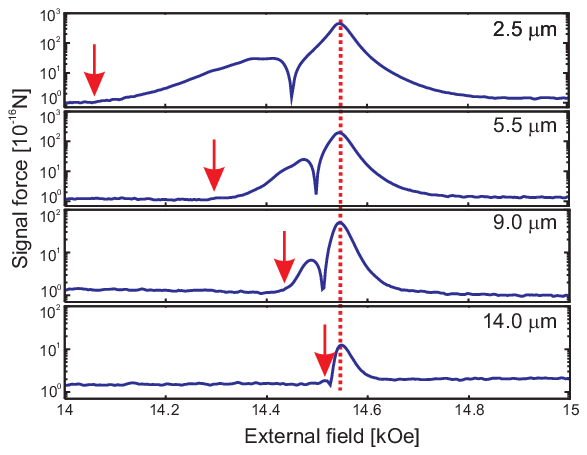}
\caption{} \label{Figure 1}
\end{figure}

\begin{figure}
\includegraphics [angle=0,width=7cm]{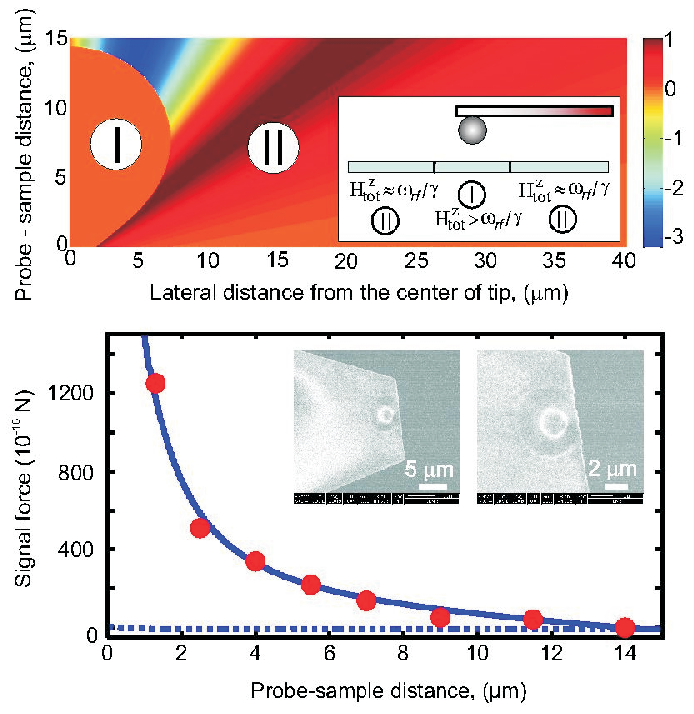}
\caption{} \label{Figure 2}
\end{figure}

The cantilever is mounted on top of a double scanning stage,
comprised of a 3D Attocube scanner \cite{Attocube} for coarse
scanning and a piezotube for fine scanning. The optical feedback
control of the Attocube allows to position and move the cantilever
stage with an accuracy better than 250 nm. The microwave power is
used to manipulate the sample magnetization and is generated by
the Giga-tronics 12000A synthesizer at a frequency $\omega_{\rm
rf}/2\pi$ = 9.55 GHz, 79 mW of power and amplitude modulated with
a modulation depth of 70\%. It is fed into a strip line resonator
with the broad resonant characteristics and allows to record FMRFM
spectra in the frequency range between 9 and 11.5 GHz. A more
detailed description of the microscope can be found elsewhere
\cite{Nazaretski 2006}. We use a commercially available silicon
nitride cantilever with a fundamental resonant frequency
$\omega_c/2\pi$ $\approx$ 8.06 kHz and a spring constant $k \sim
10$ mN/m. The magnetic tip is a 2.4 $\mu$m diameter spherical
Nd$_2$Fe$_{14}$B particle \cite{Magnequench} shown in the inset to
Fig.~\ref{Figure 2}. We removed the original tip of the cantilever
by focused ion milling, and manually glued the magnetic sphere to
the cantilever with Stycast 1266 epoxy  in the presence of an
aligning magnetic field of a few kOe. The 50 nm thick permalloy
film was deposited on a 20 nm thick Ti adhesion layer on a 100
$\mu$m thick silicon wafer. The permalloy was capped with a
protective 20 nm Ti layer.
An approximately 2 $\times$ 2 mm$^2$ sample was glued to the strip
line resonator and the film plane was oriented perpendicular to
the direction of the external magnetic field $H_{\rm ext}$.
In Fig.~\ref{Figure 1}. we show FMRFM spectra recorded as a
function of the probe-sample distance at a constant temperature $T
= 11.000 \pm 0.005$ K.  Each spectrum displays two distinctive
features: the main resonance signal, which occurs at approximately
$H_{\rm ext}$ $\approx$ 14.54 kOe, and the secondary resonance
structure at lower fields. The structure of the FMRFM signal is
reminiscent of those observed in permalloy microstructures
\cite{Mewis 2006}. The intensities of both features decrease as
the probe magnet is moved away from the surface of the sample. It
is important to note that retraction of the probe does not change
the position of the main resonance peak significantly, and at the
same time the width of the secondary feature changes
substantially. The quality factor $Q$ of the cantilever decreases
from $\sim 11,000$ to $\sim 6,000$ as the probe approaches the
sample. The change in $Q$ is due to tip-sample interactions (other
than magnetic resonance) and is consistent with previous reports
\cite{Dorofeyev 1999, Nazaretski 2006 a}. We measured $Q$ by two
methods (ring-down technique and swept frequency through
resonance) at each probe-film spacing , and subsequently
calculated the magnitude of the force signal acting on the
cantilever, Fig. \ref{Figure 1}.

In general, the force $\mathbf F$ detected in an MRFM experiment
is a convolution of $\delta \mathbf m (\mathbf r,t)$ (the change
in sample magnetization due to {\it rf} manipulation) with the
field gradient $\nabla \mathbf H_{\rm tip}(\mathbf r) $ of the
magnetic tip. The force is given by the following volume integral:
$\mathbf F = \int_{V_s} (\delta \mathbf m (\mathbf r,t) \cdot  \bf
\nabla)\mathbf H_{\rm tip}(\mathbf r) d\mathbf r.$
In our experiments, the spherical shape of the probe magnet allows
analytical calculation of its magnetic field profile $\mathbf
H_{\rm tip}(\mathbf r)$ \cite{Jackson 1975}, and is used to
provide precise knowledge of the magnetization term $\delta
\mathbf m (\mathbf r,t)$ needed to interpret FMRFM spectra
correctly.

The total magnetic field inside the sample is $\mathbf H_{\rm tot}
= \mathbf H_{\rm ext} + \mathbf H_{\rm tip} + \mathbf H_{d}$,
where $\mathbf H_{\rm ext}$ is a uniform external magnetic field,
$\mathbf H_{\rm tip}$ is a nonuniform magnetic field of the probe
magnet and $\mathbf H_{d}$ is the demagnetizing field.
The exact spatial profile of $\mathbf H_{\rm tot}$ depends on the
total magnetic moment of the probe magnet, the probe-film spacing
and the relative orientation of $\mathbf H_{\rm ext}$ with respect
to the orientation of  probe magnet magnetization (in our case
they are parallel). The well defined shape of the magnetic tip
allows us to schematically divide the sample into two regions
according to the magnitude of the $\hat{\mathbf z}$ component of
the total magnetic field $H_{\rm tot}^z$. The first, region I, is
a circular region directly under the magnetic tip where its field
is significant and positive. The area of this region is determined
by the distance between the center of the probe tip and the
sample. In region II the field of the probe is much weaker and
negative and its area encompasses the remaining sample area of $
\sim 2\times 2 \, \rm mm^2$. The schematics of two regions is
shown in the inset to Fig. \ref{Figure 2}.

For a conventional FMR experiment the expected resonance field for
the uniform FMR mode is $H_{\rm res}^u = \frac{\omega_{\rm
rf}}{\gamma} + 4 \pi M_s$,
where we neglect the anisotropy contribution, \cite{Frait 1977,
Nazaretski 2006 a}. For our experimental parameters $ \gamma/2\pi
= 2.89 \pm 0.05 $ GHz/kOe \cite{Nazaretski 2006 a} and
$\omega_{\rm rf}/2\pi = 9.55 $ GHz, we obtain the value of $H_{\rm
res}^u = 14.6$ kOe, which agrees within the error with the
observed resonance field for the main peak (dotted line in
Fig.~\ref{Figure 1}); we thereby attribute it to the resonance
originating from region II of the sample and representing its
large area. The main resonance peak in Fig. \ref{Figure 1} can be
understood as the fundamental FMR mode observed in conventional
FMR experiments \cite{Kittel 1951, Sparks 1969}, modified by the
tip field. Analytical derivation of the exact profile of such a
modified mode is difficult, so we have performed micromagnetic
simulations based on the numerical solution of the
Landau-Lifshitz-Gilbert equation \cite{Gilbert 1955}. For
simulation we used a damping constant $\alpha$ = 0.01, an exchange
constant A = 1.4$\times$10$^{-6}$ erg$\cdot cm^{-1}$ and values of
4$\pi M_s$ = 13.2 kG for the probe magnet and 4$\pi M_s$ = 11.3 kG
for the permalloy film were measured independently by the SQUID
magnetometry \cite{Nazaretski 2006 b}. Simulations indicate that
the FMR mode excited in region II of the sample at the resonant
field $H_{\rm res}$ does not penetrate significantly into spatial
region I, where $H_{\rm tot}^z > \frac{\omega_{\rm rf}}{\gamma}$.
We will present details of the analysis elsewhere
\cite{Pelekhov...}.

We assign the lower field feature in the FMRFM spectra shown in
Fig.~\ref{Figure 1} to the resonance contributions originating
from the {\em localized} FMR excitations spatially confined
approximately to region I of the sample. In this region, the
resonance occurs at lower values of $H_{\rm ext}$ than that of the
main peak (Fig. \ref{Figure 1}, dotted line). The frequency shift
of {\em localized} FMR is determined by two factors: (1) the
strength of the tip field $H_{\rm tip}$ at the sample surface, and
(2) the effect of mode confinement to the spatial region I with
characteristic dimensions defined by the tip-sample distance,
which further increases the {\em local} mode frequency relative to
the bulk resonance by a value $\Delta \omega_{conf}(r')$.  Both
effects cause the local resonance to occur at the external field
value that is {\em lower} than that of the bulk resonance by the
amount $\Delta H_{\rm ext}\approx - H_{\rm tip}(r') -
\Delta\omega_{conf}(r')/\gamma$. From numerical simulations
\cite{Pelekhov...}, for a confinement within a disc of radius
$r'\sim 10 \mu$m, $\Delta\omega_{conf}(r')/\gamma\approx 30$ Oe,
which combined with the estimated value of $ H_{\rm tip}(r')
\approx$ 20 Oe results in a total shift consistent with
experimental findings. Both the tip field $H_{\rm tip}$  and
$\Delta \omega_{conf}(r')$ decrease as the probe magnet is
retracted away from the film surface, and local modes merge into
the main resonance (Fig.~\ref{Figure 1}, dotted line).

The non-lorentzian and broad shape of the signal possibly
indicates the presence of {\rm multiple} modes contributing to the
resonance. While numerical simulations are required to determine
the possibility of such modes in particular geometry/materials, we
believe that their appearance is generic in thin-film soft magnets
and is induced by the local field inhomogeneity. We will present
the detailed theoretical analysis elsewhere\cite{Pelekhov...}.

The normalized FMRFM spatial force map obtained from the uniform
FMR mode modified by the tip field is shown in Fig.~\ref{Figure
2}. The semicircular region of the plot where the tip-sample
interaction force is set to zero corresponds to region I. In the
lower panel of Fig.~\ref{Figure 2} we show the intensity of the
main resonance signal as a function of the tip-sample distance and
compare simulations with the experiment. The dashed line shows the
expected force signal for the uniform FMR mode, not modified by
the tip field. In this case there is no force exerted by a
uniformly magnetized infinite film on a spherical probe tip.

In conclusion, we conducted FMRFM experiments in a thin permalloy
film.  We performed quantitative analysis of the force exerted by
the fundamental mode and observed locally excited FMR. We
conducted simulations and determined two distinctive regions of
the sample contributing to the FMRFM spectra. We find clear
evidence for local modification of the FMR mode structure by the
probe tip providing insight into the interaction of the probe tip
with ferromagnetic samples in FMRFM.

The work performed at Los Alamos National Laboratory was supported
by the US Department of Energy, Center for Integrated
Nanotechnologies, contract W-7405-ENG-36 at Los Alamos National
Laboratory and contract DE-AC04-94AL85000 at Sandia National
Laboratories. The work at Ohio State University was supported by
the US Department of Energy through grant DE-FG02-03ER46054. The
work at Naval Research Laboratory (NRL) was supported by the
Office of Naval Research through the Institute for Nanoscience at
NRL.

\newpage

Figure Caption

\vspace{3cm}

Figure 1: Evolution of the FMRFM signals as a function of the
probe-film spacing. Dotted line indicates the position of the main
resonance peak, independent of the probe-film distance. Arrows
mark the onset of the lower field resonance feature. Experimental
parameters: $\omega_{\rm rf}/2\pi$=9.55 GHz, T = 11 K

\vspace{2cm}

Figure 2: Upper panel: force map of the FMRFM probe-film
interaction. The force acting on a cantilever due to an elementary
ring-shaped area is calculated as a function of radius of the ring
and the cantilever-film spacing. The force map is normalized to a
maximum positive force value at each probe-film distance. Force
contribution from region I of the sample is set to zero because
the fundamental resonant mode which occurs at $H_{\rm res}$ does
not significantly penetrate into region I where $H_{\rm tot}^z >
\frac{\omega_{\rm rf}}{\gamma}$. The inset shows schematically the
probe-sample arrangement and two regions of the sample
contributing to the FMRFM signal. Lower panel: integrated
probe-film interaction force as a function of the cantilever-film
spacing for the main resonance peak at $H_{\rm ext}$$\sim 14.54$
kOe. Solid symbols represent experimental points and solid line is
the result of calculations based on the exclusion of region I. The
dashed line shows the force if region I were included in
integration. Two insets show SEM micrographs of the cantilever
tip.

\newpage

\end{document}